\documentclass[12pt]{iopart}

\usepackage{graphicx}
\usepackage{lineno}
\begin{document}


\title[J.Z. Wang et al., VSHPIC]{VSHPIC: A Particle-In-Cell Algorithm Based On Vector Spherical Harmonics Expansion}

\author{Jianzhao Wang$^1$, Weiming An$^{1,2,^\dagger}$, Rong Tang$^1$, Weiyu Meng$^1$ \& Jiayong Zhong$^{1,2}$}

\address{$^1$ Department of Astronomy, Beijing Normal University, No.19, Xinjiekouwai St, Haidian District, Beijing 100875, China}
\address{$^2$ Institute for Frontiers in Astronomy and Astrophysics, Beijing Normal University, Changping District, Beijing 102206, China}
\footnotetext{Author of correspondence: anweiming@bnu.edu.cn}
\vspace{10pt}
\begin{indented}
\item[]January 2024
\end{indented}

\begin{abstract}
The Particle-in-Cell (PIC) simulation has been a widely used method for studying plasma physics. However, fully three-dimensional PIC simulations always require huge computational resources. For problems with near azimuthal symmetry, recent work has shown that expanding all the quantities defined on the grid in azimuthal harmonics and truncating the expansion can improve the code efficiency. In this paper, we describe a novel parallel algorithm for efficiently simulating three-dimensional near-spherical symmetry problems. Our approach expands all physical quantities in the $\theta$ and $\phi$ directions in spherical coordinates using vector spherical harmonics. The code is capable of simulating three-dimensional asymmetric scenarios by accurately tracking the evolution of distinct individual modes while preserving the charge conservation law. The fundamental dispersion relation of EM waves in the plasma has been obtained using VSHPIC simulation results. The code also shows a well strong scalability up to more than 1000 cores.

\end{abstract}

\vspace{2pc}

\section{Introduction} \label{sec:intro}

Particles-in-Cell (PIC) simulation is one of the important methods for studying the collective behavior of a plasma. It can self-consistently track the plasma particles motion in both self and external electromagnetic (EM) fields. The PIC algorithm can also be parallelized and scale to more than 1 million processors on super computers\cite{Fonseca_2013,VINCENTI201822}. In the fields such as beam-plasma interaction, laser-plasma interaction and astrophysics, the PIC simulations can provide insights and information that complement the information obtained through traditional experimental and theoretical approaches\cite{RevModPhys.55.403}. However, a large scale fully three-dimensional PIC simulations may cost significant computational resources. Converting a three-dimensional simulation to a one-dimensional or two-dimensional simulation is one of the common ways to reduce the computational burden. But this may result in the loss of some physical effects. Mode decomposition is a new way to reduce computational load without the loss of physical effects. A.F. Lifschitz et al. developed a PIC code in cylindrical coordinates based on a Fourier decomposition along the azimuthal direction\cite{2009JCoPh.228.1803L}. A.Davidson et al. \cite{2015JCoPh.281.1063D} improved this work by using a Marder's correction \cite{MARDER198748} to maintain the charge conservation law under the Fourier decomposition. A.Davidson's code can also be truncated at an arbitrary number of mode in the Fourier decomposition. R.Lehe et al. developed a new spectral quasi-cylindrical PIC algorithm based on Hankel transform and Fourier transform\cite{2016CoPhC.203...66L}. The code is able to avoid numerical dispersion of the EM wave. In addition, F. Li et al. combined azimuthal decomposition with a quasi-static PIC code for accelerating the simulation of plasma based accelerator\cite{2021CoPhC.26107784L,LI2022111599}. However, all these PIC codes are using cylindrical coordinates.

In this paper, we present a new parallel PIC code that uses vector spherical harmonics to expand EM fields and charge and current densities in spherical coordinates. The code is named as the Vector Spherical Harmonic PIC (VSHPIC) code. The expansion in the code can be truncated at any mode number as needed. A rigorous approach maintains the charge conservation law. Through the spherical harmonics expansion in $\theta$ and $\phi$ directions, a three-dimensional simulation can be degraded into a one-dimensional simulation along $r$ direction, which can significantly reduce the computational load.

The paper is organized as follows. Section \ref{sec:vshpic} describes each part of the VSHPIC algorithm, including the decomposition of the EM field based on vector spherical harmonics, the current deposition method, the EM interpolation and the particle pusher. In section \ref{sec:res} we show the VSHPIC simulation results of a hot plasma. At last, we summarize the work and discuss the future work about the algorithm.


 
\section{The algorithm in VSHPIC} \label{sec:vshpic}

\subsection{Fields expansion using vector spherical harmonics } 

The vector spherical harmonics (VSH) are an extension of the scalar spherical harmonics for the use with vector fields. There are different definitions for the vector spherical harmonics (VSH). We choose B.Carrascal's definition\cite{Carrascal_1991}. A vector field can be expanded into series based on VSH.  For example, the EM fields can be expanded as:
\begin{equation}
  \mathbf{E}=\sum_{l=0}^{\infty} \sum_{m=-l}^{l}\left(E_{l m}^{r} \mathbf{Y}_{l m}+E_{l m}^{(1)} \mathbf{\Psi}_{l m}+E_{l m}^{(2)} \mathbf{\Phi}_{l m}\right) \label{comE}
\end{equation}
\begin{equation}
  \mathbf{B}=\sum_{l=0}^{\infty} \sum_{m=-l}^{l}\left(B_{l m}^{r} \mathbf{Y}_{l m}+B_{l m}^{(1)} \mathbf{\Psi}_{l m}+B_{l m}^{(2)} \mathbf{\Phi}_{l m}\right) \label{comB}
\end{equation}
, where $\mathbf{Y}_{l m}$, $\mathbf{\Psi}_{l m}$ and $\mathbf{\Phi}_{l m}$ are the basis vectors for each mode ($l,m$) of the VSH (see its definition in Appendix A). The directions of three orthogonal basis vectors are labeled as $r,(1),(2)$. $E_{lm}^r$, $E_{lm}^{(1)}$ and $E_{lm}^{(2)}$ are complex amplitudes for each mode of the electric field in three directions, so are the magnetic field's complex amplitudes $B_{lm}^r$, $B_{lm}^{(1)}$ and $B_{lm}^{(2)}$. Note that the bold letters denote vectors, regular letters denote scalars in this paper.

Similarly, the current density $\mathbf{J}$ and the charge density $\mathbf{\rho}$ can be expanded as,
\begin{equation}
  \mathbf{J}=\sum_{l=0}^{\infty} \sum_{m=-l}^{l}\left(J_{l m}^{r} \mathbf{Y}_{l m}+J_{l m}^{(1)} \mathbf{\Psi}_{l m}+J_{l m}^{(2)} \mathbf{\Phi}_{l m}\right) \label{comJ}
\end{equation}
\begin{equation}
  \rho=\sum_{l=0}^{\infty} \sum_{m=-l}^{l}\rho_{lm}\,{Y}_{l m} \label{comq}
\end{equation}
, where $J_{lm}^r$, $J_{lm}^{(1)}$ and $J_{lm}^{(2)}$ are complex amplitudes for each mode of the current density. Since the charge density $\mathbf{\rho}$ is a scalar, it should be expanded with the scalar spherical harmonics $Y_{lm}$. The $\rho_{lm}$ is complex amplitude of the charge density. All the mode amplitudes for EM fields and the densities are functions of the radial position $r$ and the time $t$.

According to the divergence and the curl of the VSH (Eqs.(\ref{div1}) to (\ref{curl3})),  the divergence and the curl of the electric field are shown below,
\begin{equation}
  \nabla \cdot \mathbf{E}=\sum_{l=0}^{\infty} \sum_{m=-l}^{l}\left(\frac{\mathrm{d} E_{l m}^{r}}{\mathrm{d} r}+\frac{2}{r} E_{l m}^{r}-\frac{l(l+1)}{r} E_{l m}^{(1)}\right) Y_{l m} \label{divF}
\end{equation}
\begin{eqnarray}
  \nabla \times \mathbf{E} & = \sum_{l=0}^{\infty} \sum_{m=-l}^{l}\left(-\frac{l(l+1)}{r} E_{l m}^{(2)} \mathbf{Y}_{l m}-\left(\frac{\mathrm{d} E_{l m}^{(2)}}{\mathrm{d} r}+\frac{1}{r} E_{l m}^{(2)}\right) \mathbf{\Psi}_{l m} \right. \nonumber \\
  & \ \ \left. +\left(-\frac{1}{r} E_{l m}^{r}+\frac{\mathrm{d} E_{l m}^{(1)}}{\mathrm{d} r}+\frac{1}{r} E_{l m}^{(1)}\right) \mathbf{\Phi}_{l m}\right) \label{curlF}
\end{eqnarray}
The magnetic field has similar expressions as above.

Unless specified, henceforth we use the normalized units, in which the time is normalized to $\omega_p^{-1}$, the length is normalized to $c/\omega_p$, the mass is normalized to the electron rest mass $m_e$ and the density is normalized to the plasma density $n_p$. In normalized units, the Maxwell's equations become
\begin{eqnarray}
  & \nabla \cdot \mathbf{E}=\rho \label{max1}\\
  & \nabla \cdot \mathbf{B}=0 \label{max2}\\
  & \frac{\partial \mathbf{E}}{\partial t}=\nabla \times \mathbf{B}-\mathbf{J} \label{max3}\\
  & \frac{\partial \mathbf{B}}{\partial t}=-\nabla \times \mathbf{E} \label{max4}
\end{eqnarray}
By substituting the EM fields and current and charge densities expanded with VSH, we have the equations for each mode's amplitudes in the Maxwell's equations,
\begin{eqnarray}
  & & \frac{1}{r^2}\frac{\partial}{\partial r}\left(r^2E_{l m}^{r}\right)-\frac{l(l+1)}{r} E_{lm}^{(1)}=\rho_{lm} \label{Max1}\\
  & & \frac{1}{r^2}\frac{\partial}{\partial r}\left(r^2B_{l m}^{r}\right)-\frac{l(l+1)}{r} B_{l m}^{(1)}=0 \label{Max2}\\
  & & \frac{\partial E_{lm}^r}{\partial t}=-\frac{l(l+1)}{r} B_{lm}^{(2)}-J_{lm}^{r} \label{Max3.1}\\
  & & \frac{\partial E_{lm }^{(1)}}{\partial t}=-\frac{1}{r}\frac{\partial}{\partial r}\left(rB_{l m}^{(2)}\right)-J_{lm}^{(1)} \label{Max3.2}\\
  & & \frac{\partial E_{lm}^{(2)}}{\partial t}=-\frac{1}{r} B_{lm}^{r}+\frac{1}{r}\frac{\partial}{\partial r}\left(rB_{l m}^{(1)}\right)-J_{lm}^{(2)} \label{Max3.3}\\
  & & \frac{\partial B_{lm}^r}{\partial t}=\frac{l(l+1)}{r} E_{lm}^{(2)} \label{Max4.1}\\
  & & \frac{\partial B_{lm }^{(1)}}{\partial t}=\frac{1}{r}\frac{\partial}{\partial r}\left(rE_{l m}^{(2)}\right) \label{Max4.2}\\
  & & \frac{\partial B_{lm}^{(2)}}{\partial t}=\frac{1}{r} E_{lm}^{r}- \frac{1}{r}\frac{\partial}{\partial r}\left(rE_{l m}^{(1)}\right) \label{Max4.3}
\end{eqnarray}  

\begin{figure}[h]
\centering
\includegraphics[scale=.8]{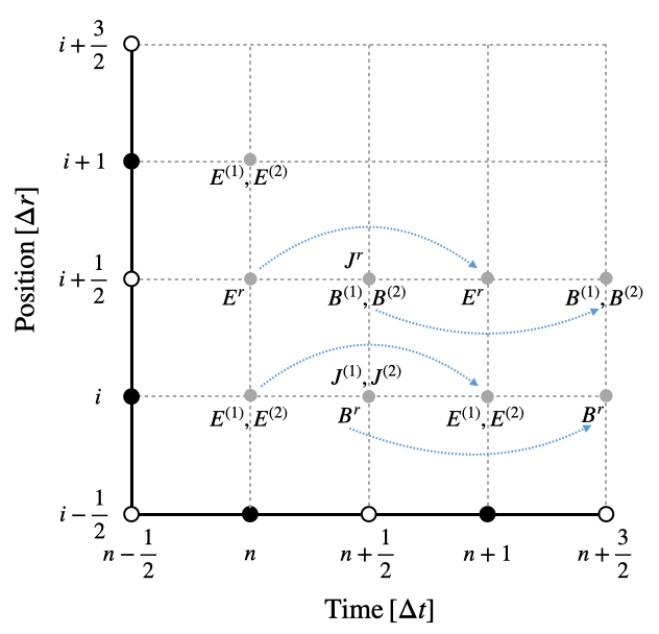}
\caption{The layout of the components of the discrete EM fields, current density and charge density on the grids of $r$ and $t$. The black dots on the axis represent the integer grid points. The white dots represent the staggered grid points. The blue dashed curve with arrows show the leap-frog process for advancing the EM fields.}
\label{fig:1}
\end{figure}

Then we can solve the above equations Eqs. (\ref{Max1}-\ref{Max4.3}) using the finite-difference in time-domain (FDTD) method. In VSHPIC, we discrete Eqs. (\ref{Max1}-\ref{Max4.3}) with central difference in $r$ and $t$, which has the second order accuracy. The layout of the EM fields and densities on the grid is shown in Figure \ref{fig:1}. Along the time axis, the electric field is defined at the integer time points $t_n=n\Delta t$ for $n=1,\cdots,N_t$, where $N_t$ is the maximum number of time steps in the simulation and $\Delta t$ is the time step. The magnetic field and the current are defined at the staggered time points $t_{n+\frac{1}{2}} = (n+\frac{1}{2})\Delta t$. Along the $r$ axis, $B^r,\,E^{(1)},\,E^{(2)},\,J^{(1)},\,J^{(2)}$ are defined on the integer grid points $r_i=i\Delta r$ for $i=1,\cdots,N_r$, while $E^r,\,B^{(1)},\,B^{(2)},\,J^{r}$ are defined on the staggered grid points $r_{i+\frac{1}{2}}=(i+\frac{1}{2})\Delta r$, where $N_r$ is the total number of grids along $r$ direction and $\Delta r$ is the grid size. As a result, we obtain the following equations for each mode in the finite-difference form,
\begin{eqnarray}
  & \frac{ r^2_{i+\frac{1}{2}} E^{r,n}_{lm,i+\frac{1}{2}} - r^2_{i-\frac{1}{2}} E^{r,n}_{lm,i-\frac{1}{2}} }{r_i^2 \Delta r} - \frac{ l(l+1)E^{(1),n}_{lm,i} }{r_i} = \rho_{lm,i}^n \label{fdtd1}\\
  & \frac{ r^2_{i+1} B^{r,n+\frac{1}{2}}_{lm,i+1} - r^2_{i} B^{r,n+\frac{1}{2}}_{lm,i} }{r_{i+\frac{1}{2}}^2 \Delta r} - \frac{ l(l+1)B^{(1),n+\frac{1}{2}}_{lm,i+\frac{1}{2}} }{r_{i+\frac{1}{2}}} = 0 \label{fdtd2}\\
  & \frac{E_{l m, i+\frac{1}{2}}^{r, n+1}-E_{l m, i+\frac{1}{2}}^{r, n}}{\Delta t}= -\frac{l(l+1)}{r_{i+\frac{1}{2}} } B_{l m, i+\frac{1}{2}}^{(2), n+\frac{1}{2}} -J_{l m, i+\frac{1}{2}}^{r, n+\frac{1}{2}}\label{fdtd3.1}\\
  & \frac{E_{l m, i}^{(1), n+1}-E_{l m, i}^{(1), n}}{\Delta t}= -\frac{r_{i+\frac{1}{2}} B_{l m, i+\frac{1}{2}}^{(2), n+\frac{1}{2}}-r_{i-\frac{1}{2}} B_{l m, i-\frac{1}{2}}^{(2), n+\frac{1}{2}}}{r_i\Delta r} -J_{l m, i}^{(1), n+\frac{1}{2}} \label{fdtd3.2}\\
  & \frac{E_{l m, i}^{(2), n+1}-E_{l m, i}^{(2), n}}{\Delta t}= -\frac{1}{r_i } B_{l m, i}^{r, n+\frac{1}{2}}+ \frac{r_{i+\frac{1}{2}} B_{l m, i+\frac{1}{2}}^{(1), n+\frac{1}{2}}-r_{i-\frac{1}{2}} B_{l m, i-\frac{1}{2}}^{(1), n+\frac{1}{2}}}{r_i\Delta r} \nonumber \\
  & \ \ \ \ \ \ \ \ \ \ \ \ \ \ \ \ \ \ \ \ \ \ \ \ \  -J_{l m, i}^{(2), n+\frac{1}{2}} \label{fdtd3.3}\\
  & \frac{B_{l m, i}^{r, n+\frac{3}{2}}-B_{l m, i}^{r, n+\frac{1}{2}}}{\Delta t}= \frac{l(l+1)}{r_i} E_{l m, i}^{(2), n+1} \label{fdtd4.1}\\
  & \frac{B_{l m, i+\frac{1}{2}}^{(1), n+\frac{3}{2}}-B_{l m, i+\frac{1}{2}}^{(1), n+\frac{1}{2}}}{\Delta t}= \frac{r_{i+1} E_{l m, i+1}^{(2), n+1}-r_i E_{l m, i}^{(2), n+1}}{r_{i+\frac{1}{2}}\Delta r} \label{fdtd4.2}\\
  & \frac{B_{l m, i+\frac{1}{2}}^{(2), n+\frac{3}{2}}-B_{l m, i+\frac{1}{2}}^{(2), n+\frac{1}{2}}}{\Delta t}= \frac{1}{r_{i+\frac{1}{2}}} E_{l m, i+\frac{1}{2}}^{r, n+1}-\frac{r_{i+1} E_{l m, i+1}^{(1), n+1}-r_i E_{l m, i}^{(1), n+1}}{r_{i+\frac{1}{2}}\Delta r}\label{fdtd4.3}
\end{eqnarray}

With given current densities, each mode of the EM fields can be advanced according to a leap-frog scheme. As shown in Figure \ref{fig:1}, with $\mathbf{B}$ and $\mathbf{J}$ located at $n+\frac{1}{2}$, we can advance $\mathbf{E}$ from $n$ to $n+1$ with Eqs. (\ref{fdtd3.1}-\ref{fdtd3.3}). By using Eqs. (\ref{fdtd4.1}-\ref{fdtd4.3}) $\mathbf{B}$ located at $n+\frac{1}{2}$ can be subsequently advanced to $n+\frac{3}{2}$ with the newly obtained $\mathbf{E}$ at $n+\frac{3}{2}$. Next, the the loop will be repeated to obtain $\mathbf{E}$ and $\mathbf{B}$ in next time step. The real and imaginary parts of the EM fields for each mode are calculated, stored and dumped into files separately in VSHPIC. Each mode of the EM fields will be combined in a post process for the purpose of visualization in three-dimensional space.


\subsection{Current deposition and correction}

From the VSH decomposition of the current density (Eq.(\ref{comJ})) and charge density (i.e. Eq. (\ref{comq})), we can obtain
\begin{eqnarray}
  & & J_{lm}^r = \int \mathbf{J}\cdot \mathbf{Y}_{lm}^*\,\mathrm{d}\Omega \label{jr}\\
  & &J_{lm}^{(1)} = \frac{1}{l(l+1)}\int \mathbf{J}\cdot \mathbf{\Psi}_{lm}^*\,\mathrm{d}\Omega \label{j1}\\
  & &J_{lm}^{(2)} = \frac{1}{l(l+1)}\int \mathbf{J}\cdot \mathbf{\Phi}_{lm}^*\,\mathrm{d}\Omega \label{j2}\\
  & &\rho_{lm} = \int \rho\cdot Y_{lm}^*\,\mathrm{d}\Omega \label{rho}
  \end{eqnarray}
In VSHPIC, each particle has the following particle shape,
\begin{eqnarray}
  & &S_p(\vec{r}-\vec{r}_p) = \frac{S\left(r-r_p\right)}{r^2} \frac{1}{\sin \theta} \delta\left(\theta-\theta_p\right) \delta\left(\phi-\phi_p\right)
\end{eqnarray}
, where $S(r-r_p)$ is the shape function in $r$ direction, $(r_p, \theta_p,\phi_p)$ is the particle position in spherical coordinates. The particle has shapes of the $\delta$ function in $\theta$ and $\phi$ directions. Therefore, the charge density and current density at $(r_i,\theta,\phi)$ for a single particle located at $(r_p, \theta_p,\phi_p)$ is,
\begin{eqnarray}
  & &\rho_i=q \frac{S\left(r_i-r_p\right)}{r_i^2} \frac{1}{\sin \theta} \delta\left(\theta-\theta_p\right) \delta\left(\phi-\phi_p\right) \\
  & &\mathbf{J}_i=q \mathbf{v}_p \frac{S\left(r_i-r_p\right)}{r_i^2} \frac{1}{\sin \theta} \delta\left(\theta-\theta_p\right) \delta\left(\phi-\phi_p\right)
\end{eqnarray}
, where $r_i$ is the position of the $i$th grid point in r direction, $q$ is the charge of a particle, $\mathbf{v}_p = (v_r, v_\theta, v_\phi)$ is the velocity of the particle. By substituting $\rho_i$ and $\mathbf{J}_i$ into Eqs. (\ref{jr} - \ref{rho}), we can obtain the charge density and current density at grid point $r_i$ for each VSH mode as follows,
\begin{eqnarray}
  \rho_{lm,i} & = \int\rho_i\cdot Y^*_{lm}\mathrm d \Omega = \rho_{00,i} \cdot P_{l}^{m}\left(\cos \theta_{p}\right) e^{-i m \phi_{p}} \label{rho00}\\
  J_{lm,i+\frac{1}{2}}^{r} & =  \int \mathbf{J}_{i+\frac{1}{2}}\cdot \mathbf{Y}^*_{lm}\mathrm d \Omega = J_{00,i+\frac{1}{2}}^{r} \cdot P_{l}^{m}\left(\cos \theta_{p}\right) e^{-i m \phi_{p}} \label{jr00}\\
  J_{lm,i}^{(1)} & = \frac{1}{l(l+1)} \int \mathbf{J}_{i}\cdot \mathbf{\Psi}^*_{lm}\mathrm d \Omega = \frac{1}{l(l+1)} \int\left(J_{\phi} \frac{1}{\sin \theta} \frac{\partial Y_{l m}^{*}}{\partial \phi}+J_{\theta} \frac{\partial Y_{l m}^{*}}{\partial \theta}\right) \mathrm{d} \Omega \nonumber \\
  & =\frac{1}{l(l+1)}\left[\frac{q v_{\phi} a}{r_i^{2}} S\left(r_i-r_{p}\right) \frac{-i m}{\sin \theta_{p}} P_{l}^{m}\left(\cos \theta_{p}\right) \mathrm{e}^{-i m \phi_{p}} \right. \nonumber \\
  & \left.+\left.\frac{q v_{\theta} a}{r_i^{2}} S\left(r_i-r_{p}\right) \frac{\mathrm{d} P_{l}^{m}(\cos \theta)}{\mathrm{d} \theta}\right|_{\theta=\theta_{p}} \mathrm{e}^{-i m \phi_{p}}\right] \\
  J_{lm,i}^{(2)} & = \frac{1}{l(l+1)} \int \mathbf{J}_{i}\cdot \mathbf{\Phi}^*_{lm}\mathrm d \Omega = \frac{1}{l(l+1)} \int\left(J_{\phi} \frac{\partial Y_{l m}^{*}}{\partial \theta}-J_{\theta} \frac{1}{\sin \theta} \frac{\partial Y_{l m}^{*}}{\partial \phi}\right) \mathrm{d} \Omega  \nonumber \\
  & =\frac{1}{l(l+1)}\left[\left.\frac{q v_{\phi} a}{r_i^{2}} S\left(r_i-r_{p}\right) \frac{\mathrm{d} P_{l}^{m}(\cos \theta)}{\mathrm{d} \theta}\right|_{\theta=\theta_{p}} \mathrm{e}^{-i m \phi_{p}} \right. \nonumber \\
  & \left.+\frac{q v_{\theta} a}{r_i^{2}} S\left(r_i-r_{p}\right) \frac{i m}{\sin \theta_{p}} P_{l}^{m}\left(\cos \theta_{p}\right) \mathrm{e}^{-i m \phi_{p}}\right] \label{j2lm}
\end{eqnarray}
, where $\rho_{00,i}=\frac{q a}{r_i^{2}} S\left(r_i-r_{p}\right)$ is the complex amplitude of the charge density for the $m=0,\,l=0$ mode, $J^r_{00,i+\frac{1}{2}}=\frac{q av_r}{r_{i+\frac{1}{2}}^{2}} S\left(r_{i+\frac{1}{2}}-r_{p}\right)$ is the complex amplitude of the current density for the $m=0,\,l=0$ mode, $a=\sqrt{\frac{(2 l+1)(l-m) !}{4 \pi(l+m) !}}$ is the normalization factor, $P_l^m$ is the associated Legendre functions. 

On the other hand, the charge density and the current density should satisfy the continuity equation in the finite difference form shown as follows,
\begin{equation}
\frac{\rho_{lm,i}^{n+1}-\rho_{lm,i}^n}{\Delta t} + \frac{r^2_{i+\frac{1}{2}} J^{r,n+\frac{1}{2}}_{lm,i+\frac{1}{2}} - r^2_{i-\frac{1}{2}} J^{r,n+\frac{1}{2}}_{lm,i-\frac{1}{2}}}{r^2_i\Delta r} - \frac{l(l+1)}{r_i}J^{(1),n+\frac{1}{2}}_{lm,i} = 0 \label{contieq}
\end{equation}
The above equation can be derived by taking the finite difference in time on both sides of the Gauss's law Eq.(\ref{fdtd1}) and using the Eq.(\ref{fdtd3.1}) and Eq.(\ref{fdtd3.2}) to cancel the E field in the derived equation. However, the continuity equation Eq.(\ref{contieq}) cannot be satisfied when we use the charge density and current density calculated from Eqs.(\ref{jr} - \ref{rho}). This is caused by that the charge density and current density are staggered in both space and time.
In order to satisfy the continuity equation (i.e. the charge conservation law) above, we can first let $l=0$ in Eq.(\ref{contieq}), then we get
\begin{equation}
\frac{\rho_{00,i}^{n+1}-\rho_{00,i}^n}{\Delta t} + \frac{r^2_{i+\frac{1}{2}} J^{r,n+\frac{1}{2}}_{00,i+\frac{1}{2}} - r^2_{i-\frac{1}{2}} J^{r,n+\frac{1}{2}}_{00,i-\frac{1}{2}}}{r^2_i\Delta r} = 0 \label{rhojr}
\end{equation}
In order to satisfy the above equation, we need to deposit the current $J^{r,n+\frac{1}{2}}_{00,i+\frac{1}{2}}$ by using the method described in reference \cite{VILLASENOR1992306} instead of using $J^r_{00,i+\frac{1}{2}}=\frac{q av_r}{r_{i+\frac{1}{2}}^{2}} S\left(r_{i+\frac{1}{2}}-r_{p}\right)$. For the continuity equation with nonzero $l$, we can substitute Eqs. (\ref{rho00}) and (\ref{jr00}) into it and take account of Eq. (\ref{rhojr}). Thus a new equation for calculating $J^{(1)}$ can be obtained,
\begin{eqnarray}
\fl  J^{(1),n+\frac{1}{2}}_{lm,i} & =& \frac{r_i}{l(l+1)\Delta t}\left[ \rho_{00,i}^{n+1} \left( P_l^m(\cos{\theta_p^{n+1}})e^{-im\phi_p^{n+1}} - P_l^m(\cos{\theta_p^{n+\frac{1}{2}}})e^{-im\phi_p^{n+\frac{1}{2}}} \right)\right. \nonumber \\
   & & \ \ -\left.\rho_{00,i}^{n} \left( P_l^m(\cos{\theta_p^{n}})e^{-im\phi_p^{n}} - P_l^m(\cos{\theta_p^{n+\frac{1}{2}}})e^{-im\phi_p^{n+\frac{1}{2}}} \right)\right] \label{j1cr}
\end{eqnarray}
The continuity equation can be kept when we use the new equations for calculating $J^r_{00}$ and $J^{(1)}_{lm}$. Note that Eq.(\ref{j2lm}) for calculating $J^{(2)}_{lm}$ does not change.

\subsection{Particle Pusher and EM Fields Interpolation} 

The VSHPIC code uses the relativistic Boris pusher to update the momenta and positions of the particles\cite{boris1970}. The Boris algorithm is used for solving the equations of motion for charged particles in EM fields in Cartesian coordinates. 
Therefore, we store the particle's position and velocity in Cartesian coordinates in VSHPIC. For calculating the EM fields felt by the particle, we first have to sum up each VSH mode of the EM fields according to Eq.(\ref{comE}) and Eq.(\ref{comB}), and convert them to the spherical coordinates (according to the definition of the vector spherical harmonics (Eqs.(\ref{YY}) to (\ref{PHI})) by using the following equations,
\begin{eqnarray}
  \fl E^{r}(r_p)=\sum_{l=0}^{l_\mathrm{max}} \sum_{m=-l}^{l} E_{l m}^{r}(r_p) Y_{l m}(\theta_p,\phi_p) \label{Er_sim} \\
  \fl E^{\phi}(r_p)=\sum_{l=0}^{l_\mathrm{max}} \sum_{m=-l}^{l}\left(\frac{\partial Y_{l m}(\theta,\phi)}{\partial \theta}\left|_{\theta_p,\phi_p}\right.E_{l m}^{(2)}(r_p) +\frac{1}{\sin \theta} \frac{\partial Y_{l m}(\theta,\phi)}{\partial \phi}\left|_{\theta_p,\phi_p}\right.E_{l m}^{(1)}(r_p) \right) \\
  \fl E^{\theta}(r_p)=\sum_{l=0}^{l_\mathrm{max}} \sum_{m=-l}^{l}\left(\frac{\partial Y_{l m}(\theta,\phi)}{\partial \theta}\left|_{\theta_p,\phi_p}\right.E_{l m}^{(1)}(r_p) - \frac{1}{\sin \theta} \frac{\partial Y_{l m}(\theta,\phi)}{\partial \phi}\left|_{\theta_p,\phi_p}\right. E_{l m}^{(2)}(r_p) \right)\label{E2_sim}
\end{eqnarray}
, where $l_\mathrm{max}$ is the maximum value of $l$ that is used in the simulation, $(r_p,\theta_p,\phi_p)$ are the particle's spherical coordinates that converted from its Cartesian coordinates and $E_{l m}^{r}(r_p)$, $E_{l m}^{(1)}(r_p)$ and $E_{l m}^{(2)}(r_p)$ are electric field components interpolated at the $r_p$. The magnetic field is calculated in the same way. Then we need to rotate the EM fields into the form of ($E_x, E_y, E_z$) and ($B_x, B_y, B_z$) at the particle location in order to push the particle with Boris pusher.

Because the EM fields are all real, then the amplitudes for $m<0$ modes is relevant to those of $m>0$ modes. Therefore, the above Eqs.(\ref{Er_sim} - \ref{E2_sim}) can be further simplified as (The details are given in Appendix B):
\begin{eqnarray}
  \fl E^{r} = E_{00}^r Y_{00}(\theta_p,\phi_p) + \sum_{l=1}^{\infty} \left[ E_{l0}^r Y_{l0}(\theta_p,\phi_p) + 2 \sum_{m=1}^{l} Re( E_{l m}^{r} Y_{l m}(\theta_p,\phi_p)) \right] \label{simpEr} \\
  \fl E^{\phi} = \frac{\partial Y_{00}(\theta_p,\phi_p)}{\partial \theta}E_{00}^{(2)} + \sum_{l=1}^{\infty} \left[ \frac{\partial Y_{l 0}(\theta_p,\phi_p)}{\partial \theta}E_{l 0}^{(2)} \right. \nonumber \\
  \left.+2 \sum_{m=1}^{l} Re\left(\frac{\partial Y_{l m}(\theta_p,\phi_p)}{\partial \theta}E_{l m}^{(2)} +\frac{1}{\sin \theta} \frac{\partial Y_{l m}(\theta_p,\phi_p)}{\partial \phi} E_{l m}^{(1)} \right) \right] \label{simpEph} \\
  \fl E^{\theta} = \frac{\partial Y_{00}(\theta_p,\phi_p)}{\partial \theta}E_{00}^{(1)} + \sum_{l=1}^{\infty} \left[ \frac{\partial Y_{l 0}(\theta_p,\phi_p)}{\partial \theta}E_{l 0}^{(1)} \right. \nonumber \\
  \left. + 2\sum_{m=1}^{l}Re\left(\frac{\partial Y_{l m}(\theta_p,\phi_p)}{\partial \theta}E_{l m}^{(1)} - \frac{1}{\sin \theta} \frac{\partial Y_{l m}(\theta_p,\phi_p)}{\partial \phi} E_{l m}^{(2)} \right) \right] \label{simpEth}
\end{eqnarray}
The calculation of magnetic field has the same way. 

The above equations still require to calculate the normalized spherical harmonics $Y_{lm}$ and its derivatives at arbitrary ($\theta_p, \phi_p$). Since $Y_{lm} = P_{lm}(\cos\theta_p)e^{im\phi_p}$, we need to calculated the normalized associated Legendre functions $P_{lm}(\cos\theta_p)$ and the complex exponential function $e^{im\phi_p}$. The function of $e^{im\phi_p}$ is calculated in a recursive way:
\begin{equation}
e^{i(m+1)\phi_p} = e^{im\phi_p}\cdot(\cos\phi_p+i\sin\phi_p)
\end{equation}
The $P_{lm}(\cos\theta_p)$ is also calculated through the recursive equations:
\begin{eqnarray}
  \fl P_{00}(x) = \sqrt{\frac{1}{4\pi}} \label{p00}\\
  \fl P_{mm}(x) = (-1)^m\cdot (2m-1)!!\cdot\left(\sqrt{1-x^2}\right)^m\cdot \sqrt{\frac{1}{(2m)!}\frac{2m+1}{4\pi}} \\
  \fl P_{m+1,m}(x) = x\cdot \sqrt{2m+3} \cdot P_{mm}(x) \\
  \fl P_{lm}(x)=\sqrt{\frac{2l+1}{(l+m)(l-m)}}\cdot \left[ x \cdot \sqrt{2l-1} \cdot P_{l-1,m}(x) \right. \nonumber \\
  \left.- \sqrt{\frac{(l+m-1)(l-m-1)}{2l-3}} \cdot P_{l-2,m}(x) \right], \quad l>m+1\label{plm}
\end{eqnarray}
, where $x = \cos\theta_p$. 
In order to avoid recalculation, we first calculate $P_{00}(\cos\theta_p)$. Then we keep the $m$ and calculate each $P_{lm}(\cos\theta_p)$ for consecutive $l$ till it reaches $l_\mathrm{max}$. Then we increase $m$ by 1 and repeat the previous process till $m$ reaches $l_\mathrm{max}$. For example, when $l_\mathrm{max}=3$, the order for calculating $P_{lm}(\cos\theta_p)$ is $(l,m) = \{(0,0), (1,0), (2,0), (3,0), (1,1), (2,1), (3,1), (2,2), (3,2), (3,3)\}$.  

For calculating the derivative of the normalized associated Legendre function, we can use the following recursive equations:
\begin{eqnarray}
  \fl P_{00}^{\prime }(x) = 0 \\
  \fl P_{mm}^{\prime }(x) = -\frac{m x}{1-x^{2}} P_{mm}(x) \\
  \fl P_{m+1,m}^{\prime }(x)= \sqrt{2m+3}\cdot \left( 1-\frac{m x^2}{1-x^{2}}\right) P_{mm}(x) \\
  \fl P_{lm}^{\prime}(x)=\sqrt{\frac{2l+1}{(l+m)(l-m)}}\cdot \left[\sqrt{2l-1} \cdot P_{l-1,m}(x) + x\sqrt{2l-1} \cdot P_{l-1,m}^{\prime}(x) \right. \nonumber\\
  \fl \qquad \qquad \left. - \sqrt{\frac{(l+m-1)(l-m-1)}{2l-3}} \cdot P_{l-2,m}^{\prime}(x) \right], \quad l>m+1\label{plmp}
\end{eqnarray}
These equations are obtained by directly taking the $x$ derivative of the Eqs.(\ref{p00} - \ref{plm}).
In addition, since the coefficients in Eqs.(\ref{p00} - \ref{plmp}) are only related to the value of $m$ and $l$, we can calculate all these coefficients with a given $l_\mathrm{max}$ and store them in a table in the initialization subroutine of VSHPIC. 

\begin{figure}[h]
\centering
\includegraphics[scale=.8]{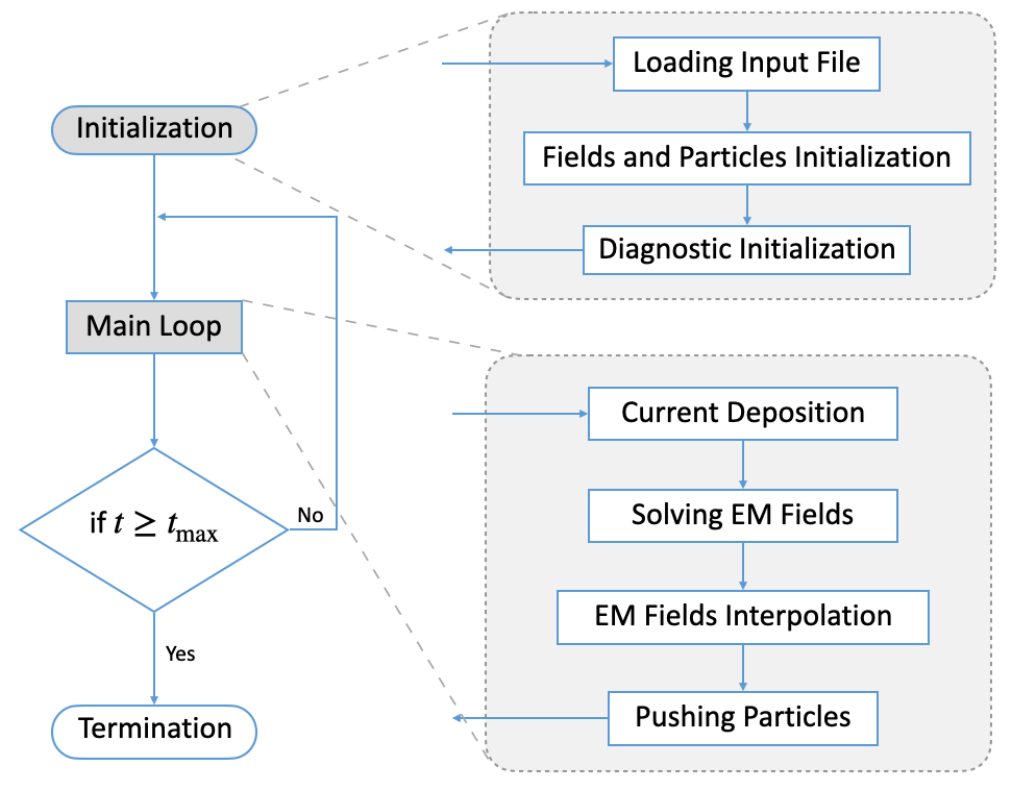}
\caption{The workflow of VSHPIC.}
\label{fig:2}
\end{figure}

\subsection{The Workflow of VSHPIC} 

The workflow of the VSHPIC algorithm is shown in Figure \ref{fig:2}. After the initialization, each mode of the current density is deposited on the grids using the Eqs.(\ref{jr00}), (\ref{j2lm}) and (\ref{j1cr}). Then each mode of the EM fields is advanced using the finite difference Eqs.(\ref{fdtd3.1} - \ref{fdtd4.3}). After obtaining the EM fields amplitudes for each mode at the new time step, they will be multiplied by the corresponding VSH and summed up at each particle's $(\theta_p, \phi_p)$. The momentum and position for the particle will be updated with Boris pusher as soon as the mode-combined EM fields are interpolated at the particle position $r_p$. The code will repeat this loop until it reaches the maximum number of time steps.

\section{The Parallelization and Simulation Results} \label{sec:res}

\subsection{Parallelization and Strong Scaling of VSHPIC}

\begin{figure}[h]
\centering
\includegraphics[scale=.6]{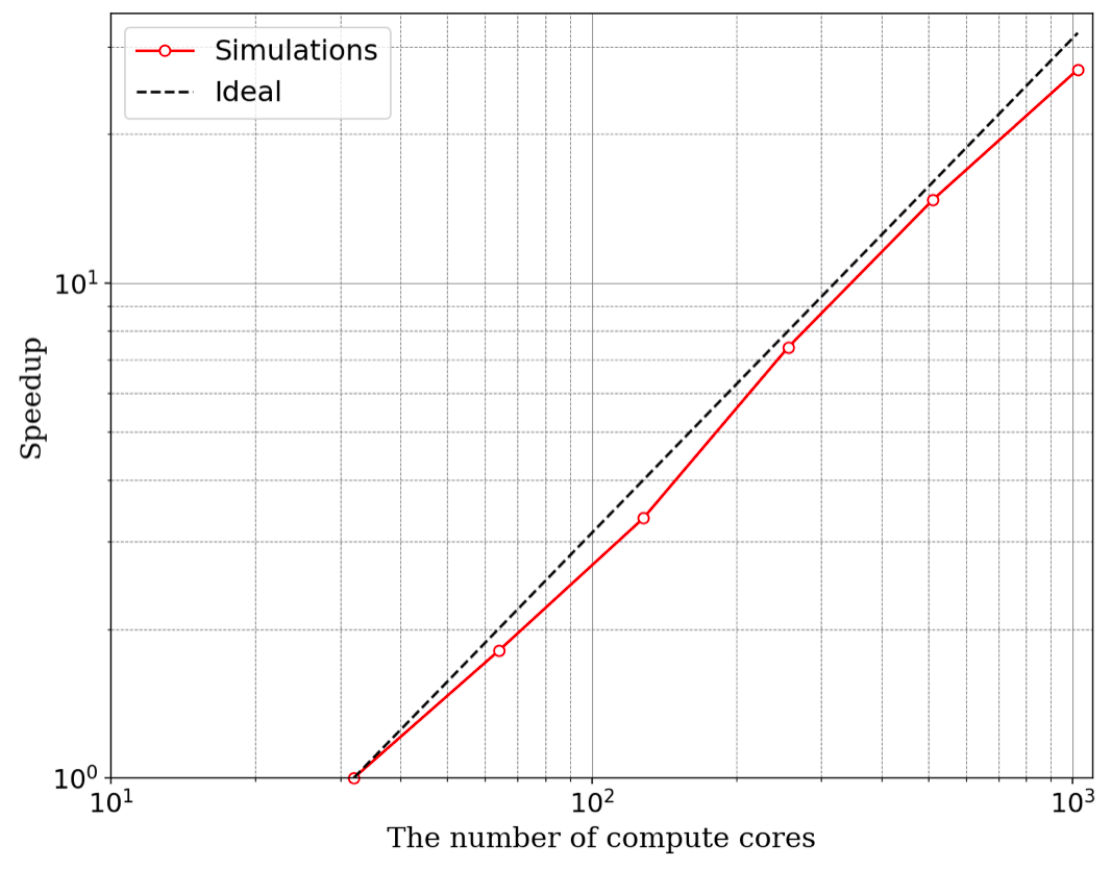}
\caption{The strong scaling benchmarks of VSHPIC}
\label{fig:3}
\end{figure}

In VSHPIC, the simulated area is uniformly divided into partitions along the radial direction. The EM fields, currents and particles in each partition will be calculated by one computing core. Copying guard cells, adding guard cells and particle moving between partitions at each time step are implemented using MPI \cite{mpi40} for the communications between computing cores. The result of the strong scaling benchmarks is presented in Fig.\ref{fig:3}. The simulations are carried out at the Partition A of Beijing Super Cloud Computing Center with the computing node of AMD EPYC 7452 32-Core Processor @ 2.35GHz. We simulate the evolution of a hot plasma with fixed ions. There are totally 8192 grid points along radial direction. We use 4 particles per cell in $r$ direction, 180 particles in $\phi$ direction and 180 particles in $\theta$ direction, giving a total of $1.06\times10^9$ particles. The modes of the EM fields and the currents are truncated at $l_\mathrm{max}=10$. The grid size is $\Delta r = 0.1$, and the time step is $\Delta t = 0.075$. The tests last for 100 time steps. In the strong scaling, we keep the parameters of the simulations and change the total number of cores. As shown in Fig.\ref{fig:3}, VSHPIC has an excellent strong scaling. In the case that using 1024 cores, the averaged computing time for pushing one particle for one time step with one core is $22.19 \mu s$.

\subsection{The Charge Conservation}

\begin{figure}[h]
\centering
\includegraphics[scale=.6]{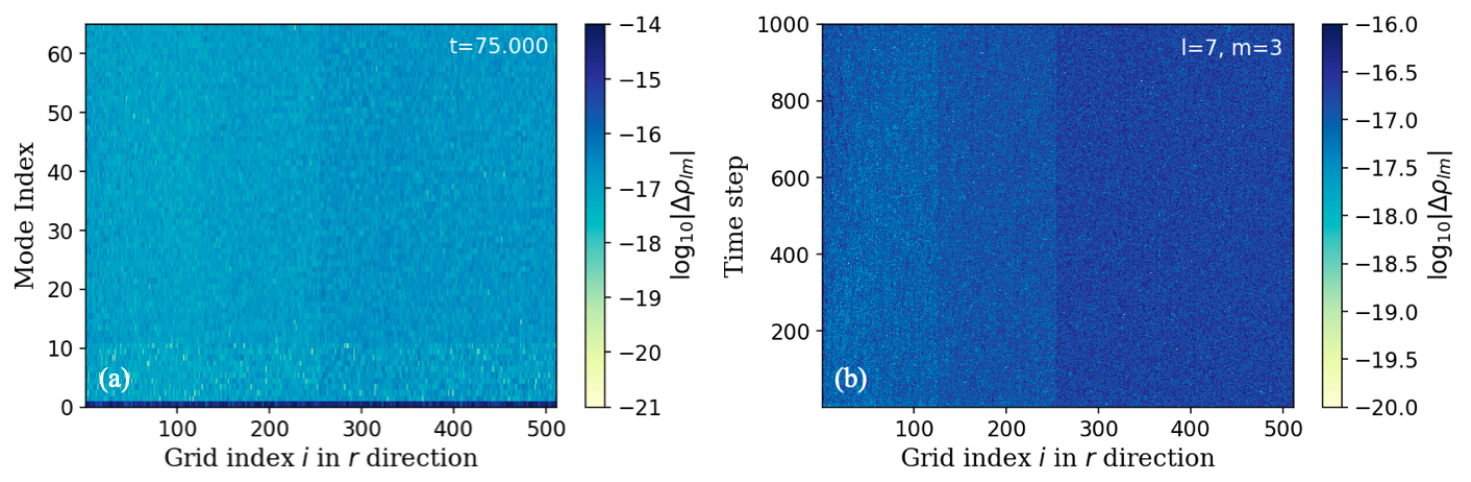}
\caption{The complex amplitude of the deviation of charge conservation (Gauss' law) in a hot plasma simulation. (a) The deviation at the 1000th time step for all the simulated modes. (b) The deviations of a single mode $l=7, m = 3$ throughout the simulation.}
\label{fig:4}
\end{figure}

As described in previous section, the charge conservation law, i.e. Eq.(\ref{contieq}), should always be satisfied in VSHPIC. In order to check that, we dump the data for each mode of $\rho$, $J^{r}$ and $J^{(1)}$ from one simulation and exam whether the value on the left hand side of Eq.(\ref{contieq}) is close to 0. We simulate the evolution of a hot plasma as a test. The electrons have thermal velocities of ($v_x=v_y=v_z=0.1c$), and the ions are fixed. 
There are totally 512 grid points along radial direction. We use 30 particles per cell in $r$ direction, 200 particles in $\phi$ direction and 30 particles in $\theta$ direction.
The modes of the EM fields and the currents are truncated at $l_\mathrm{max}=10$. The grid size $\Delta r$ is 0.1. The time step $\Delta t$ is 0.075. The simulation is run with 64 cores. The deviation of the charge conservation is calculated in the following way,
\begin{equation}
\fl \Delta \rho_{lm} = \rho_{lm,i}^{n+1}-\rho_{lm,i}^n + \frac{r^2_{i+\frac{1}{2}} J^{r,n+\frac{1}{2}}_{lm,i+\frac{1}{2}} - r^2_{i-\frac{1}{2}} J^{r,n+\frac{1}{2}}_{lm,i-\frac{1}{2}}}{r^2_i\Delta r}\Delta t - \frac{l(l+1)}{r_i}J^{(1),n+\frac{1}{2}}_{lm,i}\cdot \Delta t \label{deviation}
\end{equation}
The $\Delta \rho_{lm}$ should be a trivial number when compared with the value of $\rho_{lm}$. Note that  $\Delta \rho_{lm}$ is a complex number. We plot $\log_{10} |\Delta \rho_{lm}|$ at the 1000th time step for each mode along $r$ direction in Fig.\ref{fig:4}(a). In Fig.\ref{fig:4}(b), we plot $\log_{10} |\Delta \rho_{lm}|$ for a single mode $l=7, m = 3$ at each time step in simulation. We can see that all the deviations are smaller than $10^{-13}$, which is on or below the order of the machine error when using double precision for the floating-point numbers. The charge conservation law is well satisfied in VSHPIC.

\subsection{The Dispersion Relation of EM Wave in a Plasma}

When considering the EM wave in a plasma, the linearized fluid equation of the plasma gives,
\begin{equation}
\nabla^2 \mathbf{E} - \frac{1}{c^2}\frac{\partial^2}{\partial t^2}\mathbf{E} - \nabla(\nabla\cdot\mathbf{E}) = \frac{\omega_p^2}{c^2}\mathbf{E} \label{Ewave}
\end{equation}
With the VSH expansion of $\mathbf{E}$, i.e. Eq. (\ref{comE}), the above equation becomes
\begin{eqnarray}
  & & -\frac{l(l+1)}{r}\left( -\frac{1}{r}E_{lm}^r+\frac{1}{r}\frac{\partial}{\partial r}\left(rE_{lm}^{(1)}\right) \right) + \frac{1}{c^2}\frac{\partial^2}{\partial t^2} E_{lm}^r = -\frac{\omega_p^2}{c^2}E_{lm}^r \label{Ewave1} \\
  & & -\frac{1}{r}\frac{\partial}{\partial r}\left(-E_{lm}^r+\frac{\mathrm d}{\mathrm d r}\left(rE_{lm}^{(1)}\right) \right) + \frac{1}{c^2}\frac{\partial^2}{\partial t^2} E_{lm}^{(1)} = -\frac{\omega_p^2}{c^2}E_{lm}^{(1)} \label{Ewave2} \\
  & & \frac{l(l+1)}{r^2}E_{lm}^{(2)}- \frac{1}{r}\frac{\partial}{\partial r}\left(\frac{\partial}{\partial r}\left(rE_{lm}^{(2)}\right) \right) + \frac{1}{c^2}\frac{\partial^2}{\partial t^2} E_{lm}^{(2)} = -\frac{\omega_p^2}{c^2}E_{lm}^{(2)} \label{Ewave3}
\end{eqnarray}
Next, we focus on the $E_{lm}^{(2)}(r,t)$. In order to find its dispersion relation, we need to make two different integral transforms on $E_{lm}^{(2)}(r,t)$ in $r$ and $t$ respectively. First we will make a Fourier transform from the time domain to the $\omega$ domain. Then we need to make the spherical Bessel transform from the $r$ domain to the $k$ domain. (The details of spherical Bessel transform are given in Appendix C). The integral transforms can be expressed as follows,
\begin{eqnarray}
  & & \tilde{E}_{lm}^{(2)}(k,\omega) = \frac{1}{\sqrt{2\pi}}\int_{-\infty}^\infty \mathrm {dt} \int_0^\infty \mathrm {d r}\;  r^2j_l(kr)\exp(i\omega t)E_{lm}^{(2)}(r,t)  \label{fwdtrans}
\end{eqnarray}
, where $j_l(kr)$ is the spherical Bessel function of the first kind.Therefore, the inverse transform gives,
\begin{eqnarray}
  & & E_{lm}^{(2)}(r,t) =\sqrt{\frac{2}{\pi^3}}\int_{-\infty}^\infty \mathrm {d\omega} \int_0^\infty \mathrm {d k}\;  k^2j_l(kr)\exp(-i\omega t)\tilde{E}_{lm}^{(2)}(k,\omega)  
\end{eqnarray}
By substituting the above equation into Eq.(\ref{Ewave3}), we can obtain (The details are given in Appendix C.),
\begin{equation}
\fl \sqrt{\frac{2}{\pi^3}}\int_{-\infty}^\infty \mathrm {d\omega} \int_0^\infty \mathrm {d k}\;  k^2j_l(kr)\exp(-i\omega t)\left( k^2-\frac{\omega^2}{c^2}+\frac{\omega_p^2}{c^2} \right)\tilde{E}_{lm}^{(2)}(k,\omega)  = 0
\end{equation}
Therefore, under the high frequency assumption the $\tilde{E}_{lm}^{(2)}(k,\omega)$ has a dispersion relation $ k^2-\frac{\omega^2}{c^2}+\frac{\omega_p^2}{c^2} =0$, which is exactly the same as that of the transverse field $\mathbf{E_T}$ (which satisfies $\nabla\cdot\mathbf{E_T}=0$) in Cartesian coordinates. Note that in the space domain we make Fourier transform in Cartesian coordinates. 

\begin{figure}[h]
\centering
\includegraphics[scale=.65]{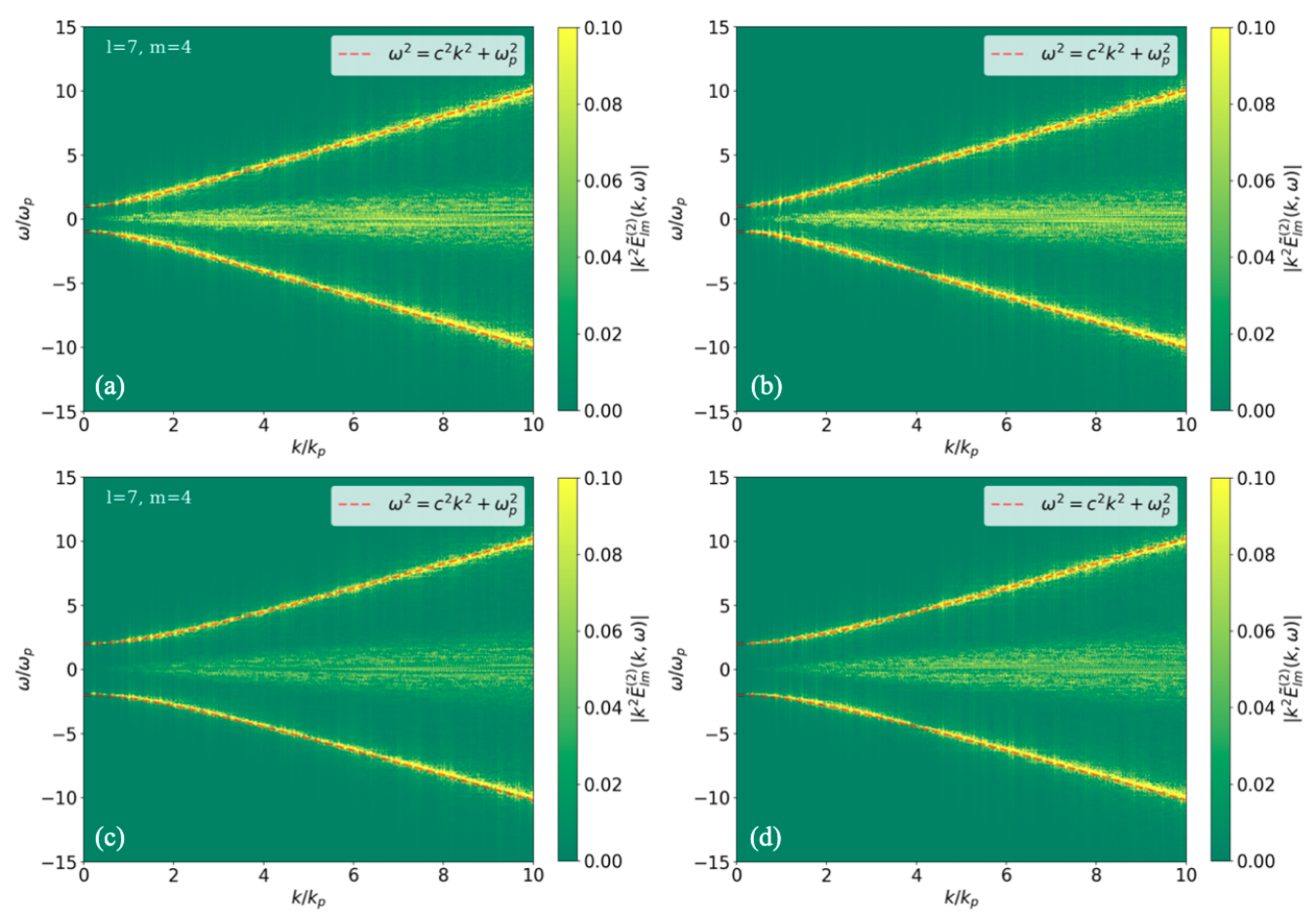}
\caption{The amplitude of $\tilde{E}_{lm}^{(2)}(k,\omega)$ that is numerically transformed from $E_{lm}^{(2)}(r,t)$ of a hot plasma simulation. (a) The mode of $l=7$ and $m=4$ with initial plasma density $n_p = 1.0$; (b) The sum of  all modes amplitudes with initial plasma density $n_p = 1.0$; (c) The mode of $l=7$ and $m=4$ with initial plasma density $n_p = 4.0$; (b) The sum of  all modes amplitudes with initial plasma density $n_p = 4.0$. The red dash line in each plot is the theoretical dispersion relation $\omega^2=c^2k^2+\omega_p^2$. In (a) and (b), $\omega_p = 1.0$. In (c) and (d), $\omega_p = 2.0$.}
\label{fig:5}
\end{figure}

To test whether VSHPIC can give a correct dispersion relation for $E_{lm}^{(2)}(r,t)$ in a plasma, we perform a hot plasma simulation (with fixed ions). In this test we have 512 grid points along radius direction. The maximum mode number is $l_\mathrm{max}=10$. We use 30 particles per cell in the $r$ direction, 80 particles in the $\theta$ and 80 particles $\phi$ direction. The grid size $\Delta r$ is 0.1. The time step $\Delta t$ is 0.075. The initial plasma density is $n_p=1.0$. Absorption boundary layers for the EM fields are applied. 
Then we numerically transform $E_{lm}^{(2)}(r,t)$ into $\tilde{E}_{lm}^{(2)}(k,\omega)$ using Eq.(\ref{fwdtrans}). The amplitude of $k^2\tilde{E}_{lm}^{(2)}(k,\omega)$ with $l=7$ and $m=4$ is plotted in Fig.\ref{fig:5}(a), while the theoretical dispersion relation $\omega^2=c^2k^2+\omega_p^2$ is plotted as the red dashed line in the same figure. We can see that the $\tilde{E}_{lm}^{(2)}(k,\omega)$ has large values on the line of the dispersion relation, which indicates that the waves that does not satisfy the dispersion relation will be evanescent. In order to show that the dispersion relation works well for all the modes, we simply sum up $|k^2\tilde{E}_{lm}^{(2)}(k,\omega)|$ for all the modes and plot it in Fig.\ref{fig:5}(b). It matches very well with the theoretical result (which is the red dashed line). We also change the initial plasma density into $n_p = 4$, where the plasma frequency $\omega_p$ becomes 2. We plot $|k^2\tilde{E}_{lm}^{(2)}(k,\omega)|$ for the mode  $l=7$ and $m=4$ and the superposition of all modes in Fig.\ref{fig:5}(c) and (d).  In this case, the cutoff frequency of $\tilde{E}_{lm}^{(2)}(k,\omega)$ becomes 2.0 as shown, which indicates a good agreement with the theoretical results. Note that the $\tilde{E}_{lm}^{(2)}(k,\omega)$ has some low frequency signals since it is not a pure transverse field, i.e. $\nabla\cdot (E_{lm}^{(2)}(r,t)\mathbf{\Phi}_{lm})\neq0$.

\section{Conclusions} \label{sec:con}

In this paper, we present a new parallel PIC code VSHPIC for efficient computation based on mode decomposition in spherical coordinates. The EM fields, current, and charge densities are expanded using vector spherical harmonics in the $\theta$ and $\phi$ directions. A new scheme for current deposition is developed to conserve the charge at each time step. Results from a hot plasma simulation show that the deviation of the continuity equation for all modes stays at the accuracy level of double precision arithmetic ($<10^{-13}$) throughout the simulation. Besides, in VSHPIC, the algorithm for calculating the values of spherical harmonics and its derivatives are carefully designed and optimized. The mode decomposition with the vector spherical harmonics results in the reduction from a three-dimensional PIC code to a one-dimensional PIC code, which can significantly decrease the computational burden compared with a conventional PIC as fewer simulated particles are required for the same physical problem with the same resolution. VSHPIC is parallelized in r direction using MPI. It shows a good strong scaling up to more than 1000 cores. A hot plasma simulation shows that the dispersion relation for the $E_{lm}^{(2)}(r)$ agrees very well with the theoretical dispersion relation. There are several directions for future work, which include further parallelization for scaling to more cores, adding modules such as particle injection, radiation, and lasers, developing the boundary condition at $r=0$. With these enhancements, we aim to apply this new PIC code in studying laser plasma interactions and astrophysics problems.

\section*{Acknowledgments}
This work is supported by the National Key R\&D Program of China 2022YFA1603200 and 2022YFA1603203, the National Natural Science Foundation of China 12075030, 12135001, 12175018 and 12325305, the Strategic Priority Research Program of the Chinese Academy of Sciences XDA25030700, the Youth Interdisciplinary Team JCTD-2022-05 and Beijing Normal University Scientific Research Initiation Fund for Introducing Talents No. 310432104 and No. 312200502503.

\section*{References}
\bibliography{iopart-num}

\clearpage

\appendix

\section{The basic formulas of vector spherical harmonics}
\setcounter{equation}{0}
\renewcommand\theequation{A.\arabic{equation}}

In this appendix, we show the definition of the vector spherical harmonics, the gradient of the scalar spherical harmonics, the divergence and the curl of the vector spherical harmonics. 

The three fundamental vector spherical harmonics (VSH) can be defined as follows \cite{RGBarrera_1985,Carrascal_1991}:
\begin{eqnarray}
  & &\mathbf{Y}_{l m}(\theta,\phi) = Y_{l m}(\theta,\phi)\  \hat{\mathbf{e}}_r \label{YY}\\
  & &\mathbf{\Psi}_{l m}(\theta,\phi) =r \nabla Y_{l m}(\theta,\phi)=\frac{1}{\sin \theta} \frac{\partial Y_{lm}(\theta,\phi)}{\partial \phi}\hat{\mathbf{e}_{\phi}} + \frac{\partial Y_{lm}(\theta,\phi)}{\partial \theta}\hat{\mathbf{e}_{\theta}} \label{PSI} \\
  & &\mathbf{\Phi}_{l m}(\theta,\phi) =\overrightarrow{\mathbf{r}} \times \nabla Y_{l m}(\theta,\phi)=\frac{\partial Y_{lm}(\theta,\phi)}{\partial \theta}\hat{\mathbf{e}_{\phi}} -\frac{1}{\sin \theta} \frac{\partial Y_{lm}(\theta,\phi)}{\partial \phi}\hat{\mathbf{e}_{\theta}} \label{PHI}
\end{eqnarray}
, where $Y_{lm}(\theta,\phi)$ is the scalar spherical harmonic of order ($l,m$), $\hat{\mathbf{e}}_r,\,\hat{\mathbf{e}_{\theta}},\,\hat{\mathbf{e}_{\theta}}$ are basis vectors in spherical coordinates, $l=0,1,2,\cdots$ and $m=-l,-l+1,\cdots,-1,0,1,\cdots,l-1,l$.

The gradient of a scalar spherical harmonics is::
\begin{equation}
  \nabla (f(r)Y_{l m})=\frac{\mathrm{d} f(r)}{\mathrm{~d} r} \mathbf{Y}_{l m}+\frac{f(r)}{r} \mathbf\Psi_{l m} \label{gradf}
\end{equation}
, where $\phi(r,\theta,\phi)=\sum_{l=0}^{\infty} \sum_{m=-l}^{l} \phi_{l m}(r) Y_{l m}(\theta, \phi)$, $\phi_{lm}(r)$ is the complex amplitude of the mode ($l,m$).

The divergence of the vector spherical harmonics is:
\begin{equation}
  \nabla \cdot\left(f(r) \mathbf{Y}_{l m}\right)=\left(\frac{\mathrm{d} f(r)}{\mathrm{d} r}+\frac{2}{r} f(r)\right) Y_{l m} \label{div1}
\end{equation}
\begin{equation}
  \nabla \cdot\left(f(r) \mathbf{\Psi}_{l m}\right)=-\frac{l(l+1)}{r} f(r) Y_{l m}
\end{equation}
\begin{equation}
  \nabla \cdot\left(f(r) \mathbf{\Phi}_{l m}\right)=0
\end{equation}
, where $f(r)$ is the complex amplitude of mode ($l,m$).

The curl of the vector spherical harmonics is:
\begin{equation}
  \nabla \times\left(f(r) \mathbf{Y}_{l m}\right)=-\frac{1}{r} f(r) \mathbf{\Phi}_{l m}
  \label{curl1}
\end{equation}
\begin{equation}
  \nabla \times\left(f(r) \mathbf{\Psi}_{l m}\right)=\left(\frac{\mathrm{d} f(r)}{\mathrm{d} r}+\frac{1}{r} f(r)\right) \mathbf{\Phi}_{l m}
  \label{curl2}
\end{equation}
\begin{equation}
  \nabla \times\left(f(r) \mathbf{\Phi}_{l m}\right)=-\frac{l(l+1)}{r} f(r) \mathbf{Y}_{l m}-\left(\frac{\mathrm{d} f(r)}{\mathrm{d} r}+\frac{1}{r} f(r)\right) \mathbf{\Psi}_{l m} \label{curl3}
\end{equation}
, where $f(r)$ is the complex amplitude of mode ($l,m$).

\section{The VSH of $(l,-m)$ modes}
\setcounter{equation}{0}
\renewcommand\theequation{B.\arabic{equation}}

For the scalar spherical harmonics with $(l,-m)$, we have
\begin{equation}
  Y_{l}^{-m}(\theta, \phi)=\sqrt{\frac{(l+m) !}{(l-m) !} \frac{2 l+1}{4 \pi}} P_{l}^{-m}(\cos \theta) e^{-i m \phi}
\end{equation}
, where $l$ and $m$ are nonnegative integers, and
\begin{equation}
  P_{l}^{-m}(\cos \theta)=(-1)^{m} \frac{(l-m) !}{(l+m) !} P_{l}^{m}(\cos \theta), \quad m=1, \ldots, l
\end{equation}
Therefore, we have
\begin{equation}
  \mathbf{Y}_{l,-m}=Y_{l,-m} \hat{\mathbf{r}}=(-1)^{m} \sqrt{\frac{(l-m) !}{(l+m) !} \frac{2 l+1}{4 \pi}} P_{l}^{m}(\cos \theta) e^{-i m \phi} \hat{\mathbf{r}} =(-1)^{m} \mathbf{Y}_{l m}^{*}
\end{equation}
Similarly, 
\begin{eqnarray}
  & &\mathbf{\Psi}_{l,-m}=(-1)^{m} \mathbf{\Psi}_{l m}^{*} \\
  & &\mathbf{\Phi}_{l,-m}=(-1)^{m} \mathbf{\Phi}_{l m}^{*}
\end{eqnarray}

For a vector field $\mathbf{E}$, it can be expanded with the VSH as,
\begin{equation}
  \mathbf{E}=\sum_{l=0}^{\infty} \sum_{m=-l}^{l}\left(E_{l m}^{r} \mathbf{Y}_{l m}+E_{l m}^{(1)} \mathbf{\Psi}_{l m}+E_{l m}^{(2)} \mathbf{\Phi}_{l m}\right) \label{vfexp}
\end{equation}
By taking the complex conjugate of the above equation, we have
\begin{eqnarray}
  \mathbf{E}^{*} & &=\sum_{l=0}^{\infty} \sum_{m=-l}^{l}\left(E_{l,-m}^{r *} \mathbf{Y}_{l,-m}^{*}+E_{l,-m}^{(1) *} \mathbf{\Psi}_{l,-m}^{*}+E_{l,-m}^{(2) *} \mathbf{\Phi}_{l,-m}^{*}\right) \nonumber \\
  & &=\sum_{l=0}^{\infty} \sum_{m=-l}^{l}(-1)^{m}\left(E_{l,-m}^{r *} \mathbf{Y}_{l m}+E_{l,-m}^{(1) *} \mathbf{\Psi}_{l m}+E_{l,-m}^{(2) *} \mathbf{\Phi}_{l m}\right)
\end{eqnarray}
When $\mathbf{E}$ is real, we have $\mathbf{E} = \mathbf{E}^*$, thus
\begin{eqnarray}
  \mathbf{E}^{*} & &=\sum_{l=0}^{\infty} \sum_{m=-l}^{l}(-1)^{m}\left(E_{l,-m}^{r *} \mathbf{Y}_{l m}+E_{l,-m}^{(1) *} \mathbf{\Psi}_{l m}+E_{l,-m}^{(2) *} \mathbf{\Phi}_{l m}\right) = \mathbf{E} \nonumber \\
& &=\sum_{l=0}^{\infty} \sum_{m=-l}^{l}\left(E_{l m}^{r} \mathbf{Y}_{l m}+E_{l m}^{(1)} \mathbf{\Psi}_{l m}+E_{l m}^{(2)} \mathbf{\Phi}_{l m}\right) 
\end{eqnarray}
Therefore, the amplitudes for each mode satisfy,
\begin{eqnarray}
  & & E_{l m}^{r}=(-1)^{m} E_{l,-m}^{r *} \\
  & & E_{l m}^{(1)}=(-1)^{m} E_{l,-m}^{(1) *} \\
  & & E_{l m}^{(2)}=(-1)^{m} E_{l,-m}^{(2) *}
\end{eqnarray}

As a result, Eq. (\ref{vfexp}) becomes
\begin{eqnarray}
  \mathbf{E} & = & E_{00}^{r} \mathbf{Y}_{00}+E_{00}^{(1)} \mathbf{\Psi}_{00}+E_{00}^{(2)} \mathbf{\Phi}_{00}+ \sum_{l=1}^{\infty} \left[ E_{l0}^{r} \mathbf{Y}_{l0}+E_{l0}^{(1)} \mathbf{\Psi}_{l0}+E_{l0}^{(2)} \mathbf{\Phi}_{l0}\right. \nonumber \\
  & &  \left.+ 2 \sum_{m=1}^{l} Re \left( E_{l m}^{r} \mathbf{Y}_{l m}+E_{l m}^{(1)} \mathbf{\Psi}_{l m}+E_{l m}^{(2)} \mathbf{\Phi}_{l m} \right) \right]
\end{eqnarray}

\section{The spherical Bessel transform}
\setcounter{equation}{0}
\renewcommand\theequation{C.\arabic{equation}}

For a function $f(r)$, where $(r \ge 0)$,  its spherical Bessel transform is defined as,
\begin{eqnarray}
  & & F(k) = \int_0^\infty r^2j_l(kr)f(r) \mathrm d r  
\end{eqnarray}
, where $j_l(kr)$ is the lth order spherical Bessel function of the first kind.
The inverse spherical Bessel transform gives,
\begin{eqnarray}
  & & f(r) =  \frac{2}{\pi} \int_0^\infty k^2F(k) j_l(kr) \mathrm d k \label{isbt}
\end{eqnarray}
, which can be easily proved by using the closure relation of the spherical Bessel function \cite{R_Mehrem_1991} as follows,
\begin{eqnarray}
   \frac{2r^2}{\pi} \int_0^\infty k^2F(k) j_l(kr)j_l(kr') \mathrm d k =\delta(r-r')
\end{eqnarray}

Next, we want to calculate the following terms by using the spherical Bessel transform,
\begin{equation}
\frac{\mathrm{d}^2f(r)}{\mathrm{d}r^2} + \frac{2}{r}\frac{\mathrm{d}f(r)}{\mathrm{d}r}-\frac{l(l+1)}{r^2}f(r)
\end{equation}
By substituting Eq. (\ref{isbt}) into the above equation, we can obtain
\begin{eqnarray}
  & &\frac{\mathrm{d}^2f(r)}{\mathrm{d}r^2} + \frac{2}{r}\frac{\mathrm{d}f(r)}{\mathrm{d}r}-\frac{l(l+1)}{r^2}f(r) \nonumber \\
  & = &\frac{2}{\pi} \int_0^\infty k^2F(k) \left(\frac{\mathrm{d}^2j_l(kr)}{\mathrm{d}r^2} + \frac{2}{r}\frac{\mathrm{d}j_l(kr)}{\mathrm{d}r}-\frac{l(l+1)}{r^2}j_l(kr)\right) \mathrm d k  \label{sbeq}
\end{eqnarray}
Since $j_l(kr)$ satisifies the $l$th order spherical Bessel equation, we have
\begin{equation}
\frac{\mathrm d ^2 j_l(kr)}{\mathrm d r^2} + \frac{2}{r}\frac{\mathrm d j_l(kr)}{\mathrm d r} + \left[ k^2 - \frac{l(l+1)}{r^2} \right] j_l(kr) = 0
\end{equation}
Thus,
\begin{equation}
\frac{\mathrm d ^2 j_l(kr)}{\mathrm d r^2} + \frac{2}{r}\frac{\mathrm d j_l(kr)}{\mathrm d r} - \frac{l(l+1)}{r^2} j_l(kr) = - k^2j_l(kr)
\end{equation}
Therefore, Eq. (\ref{sbeq}) becomes
\begin{eqnarray}
  \frac{\mathrm{d}^2f(r)}{\mathrm{d}r^2} + \frac{2}{r}\frac{\mathrm{d}f(r)}{\mathrm{d}r}-\frac{l(l+1)}{r^2}f(r) & = -&\frac{2}{\pi} \int_0^\infty k^4F(k)j_l(kr) \mathrm d k  \label{sbeqn}
\end{eqnarray}
For $E_{lm}^{(2)}(r,t)$ that satisfies Eq. (\ref{Ewave3}), we have
\begin{eqnarray}
  & & \frac{l(l+1)}{r^2}E_{lm}^{(2)}- \frac{2}{r}\frac{\partial E_{lm}^{(2)}}{\partial r} -\frac{\partial^2E_{lm}^{(2)}}{\partial r^2} + \frac{1}{c^2}\frac{\partial^2}{\partial t^2} E_{lm}^{(2)} = -\frac{\omega_p^2}{c^2}E_{lm}^{(2)} \label{Ewave3new}
\end{eqnarray}
By applying the Fourier transform in $t$ and spherical Bessel transform in $r$ for $E_{lm}^{(2)}(r,t)$, it becomes
\begin{eqnarray}
  & & \tilde{E}_{lm}^{(2)}(k,\omega) = \frac{1}{\sqrt{2\pi}}\int_{-\infty}^\infty \mathrm {dt} \int_0^\infty \mathrm {d r}\;  r^2j_l(kr)\exp(i\omega t)E_{lm}^{(2)}(r,t) 
\end{eqnarray}
, and the inverse transform gives,
\begin{eqnarray}
  & & E_{lm}^{(2)}(r,t) =\sqrt{\frac{2}{\pi^3}}\int_{-\infty}^\infty \mathrm {d\omega} \int_0^\infty \mathrm {d k}\;  k^2j_l(kr)\exp(-i\omega t)\tilde{E}_{lm}^{(2)}(k,\omega)  
\end{eqnarray}
By substituting the above equation into Eq. (\ref{Ewave3new}) and applying the Eq. (\ref{sbeqn}), we can obtain that
\begin{equation}
\fl \sqrt{\frac{2}{\pi^3}}\int_{-\infty}^\infty \mathrm {d\omega} \int_0^\infty \mathrm {d k}\;  k^2j_l(kr)\exp(-i\omega t)\left( k^2-\frac{\omega^2}{c^2}+\frac{\omega_p^2}{c^2} \right)\tilde{E}_{lm}^{(2)}(k,\omega)  = 0
\end{equation}
\end{document}